\documentclass{ws-p9-75x6-50}

\begin{document}

\title{Effect of Coriolis force on Accretion Flows around Rotating Compact Object}

\author{A. R. Prasanna \& Banibrata Mukhopadhyay}

\address{Theoretical Physics Group, Physical Research Laboratory,
Navrangpura, Ahmedabad-380009, India\\
}

\maketitle

\abstracts{
Using the generalised set of fluid equations that include the
`Coriolis force' along with the centrifugal and pressure gradient
forces, we have reanalysed the class of self similar solutions,
with the pseudo-Newtonian potential. We find that the class of
solutions is well behaved for almost the entire parameter space
except for a few selected combinations of $\gamma$ and $\alpha$ for the
co-rotating flow. The analysis of the Bernoulli number shows that
whereas it remains positive for co-rotating flow for $f>1/3$, for the
counter-rotating flow it does admit both positive and negative values,
indicating the possibility of energy transfer in either direction.
}

\section{Introduction}

One of the leading problems, pursued intensely over the last three
decades, in connection with high energy cosmic sources is the topic
of accretion disks around compact objects. There have been several
reviews\cite{nmq,kfm} on the topic covering
both theoretical and observational findings, but yet the subject is alive
with several interesting and unsolved problems. In the last decade
the theory of advection dominated accretion flows drew a lot of attention
\cite{ny,ackl,caln} and
this approach is now enlarging its scope to include convection too in the
equilibrium solutions\cite{ica,nia}.
As is well known, the class of solution used for these discussion have
been of `Self Similar' nature, which looks quite promising in explaining
several observational features\cite{emn,nbm,nmgpg}.
Though the classical solution of Narayan \& Yi\cite{ny}
has provided good insight into the properties of accretion flows, the solution
has a restriction in the parameter space with the possible breakdown
of self similarity for certain combination of the physical parameters,
$\gamma$ (the gas constant) and $\alpha$ (the viscosity parameter). This
was found recently by Bhatt \& Prasanna\cite{bp} through an analysis of the
solution perturbatively, using the well known pseudo-Newtonian potential\cite{pw}.
They showed that though self similarity
could be an excellent approximation over a large range of viscosity
parameters, for certain combination ($\alpha=0.3$, $\gamma=1.55$), the
perturbation develops a singularity. In fact it is well known that
most of the discussions, particularly in the context of modelling
accretion around black holes and neutron stars\cite{nia,c96,mn},
assume the value of gas constant $\gamma$ to range between $4/3$ to $5/3$.
Since it was shown that a singularity could develop for a value in between,
it is indeed necessary to find solutions without the pathology at least
in the parameter regime of interest.
 
It is also known that almost all the discussions, particularly in the
Newtonian regime, consider the equations of motion for the fluid in a
background geometry where the effects of rotation of the central object is
not considered. In the case of black hole accretion, a correct model
should be fully general relativistic, which if considered on a Kerr
background geometry would automatically take into effect the rotation
of the black hole through `inertial frame dragging' effect. If one wants to
create a similar scenario in a purely Newtonian discussion the only way
would be to bring into focus the `Coriolis force', wherein one writes
the equation of motion in a rotating frame. In fact the concept of
reintroducing the language of ``inertial forces" in
general relativity, though not very popular, has been found to be
useful in explaining some already known results more satisfactorily
\cite{a93,ap,m92,p97}. Particularly the fact that the ``Centrifugal force" reverses
sign very close to the compact object, could be a plausible
reason for some of the known effects. As
was shown by Abramowicz et al.\cite{anw} it is indeed possible to
introduce the concept of inertial forces in a covariant formalism that
is valid in general spacetimes, without any particular symmetry requirement.
Of these forces, the ``Coriolis" force is generally assumed to be of less
relevance in most of the Newtonian physics. However, in
general relativity the term which gives rise to the Coriolis type
field is the well known ``Lense-Thirring" effect of dragging of
inertial frames\cite{acl}, which couples the angular
momentum of the gravitating source with the particle dynamics, through
the spacetime curvature. It is indeed interesting to note that
a particle in circular orbit around a Kerr black hole at a radius
wherein the centrifugal force is zero, has to change its angular velocity
depending upon the Kerr parameter `a' to be in equilibrium. In fact
while a co-rotating particle has to decrease its angular velocity
slightly with increase in `a', the counter-rotating ones have to
increase it substantially to keep in equilibrium\cite{p97,p01}.
Though this is purely a general relativistic effect, which happens very
close to black hole, one wonders whether in principle there could be
effects of the rotation of the central gravitating source on the particles
or fluids at a distance, even in a purely Newtonian approach. It
is quite well known that while discussing disks in binary systems and
planetary rings one does consider the equations of motion in a rotating
coordinate system, that includes the Coriolis and centrifugal terms
\cite{p85,fg}.
 
With this background we now look for possible influence of Coriolis
type terms on the fluid flow in accretion disks around compact rotating objects,
particularly for the type wherein the fluid disk is in contact with the
central body. The present aim of this discussion is to find out whether
the introduction of the rotational effects through `Coriolis' type terms
in the equations of motion would alter the situation regarding the
occurrence of singularity in the perturbation solution. Further, as will be
shown, this treatment also suggests a way to distinguish between the
co and counter rotating accretion flow onto compact objects.

\section{Formalism}

As is usual in accretion disk theory we consider the height
integrated set of equations describing the steady state, axisymmetric
fluid distribution that allows one to describe all physical quantities as
functions of the cylindrical radius $R$ only.
 
Mass conservation equation yields
   \begin{equation}
      -4\pi RH\rho V ={\dot M} = {\rm constant}\,,
   \end{equation}
wherein $H=\sqrt{5/2} (c_s/\Omega_K)$, is the vertical half thickness
with $c_s$ and $\Omega_K$ being the isothermal sound speed and the
Keplerian angular velocity respectively.
 
The equation of motion for the fluid in the gravitational field of
a slowly rotating compact object is given by
   \begin{equation}
   \frac{d {\overrightarrow V}}{dt} + 2{\overrightarrow \omega} \times
{\overrightarrow V} + {\overrightarrow \omega} \times
{\overrightarrow \omega} \times {\overrightarrow R}
=  - \frac{\nabla p}{\rho} + {\overrightarrow F_g} +
\nu \nabla^2 {\overrightarrow V}
   \end{equation}
wherein, ${\overrightarrow \omega}$ is the intrinsic angular
velocity of the central
body, ${\overrightarrow V}$ the 3-velocity of the fluid element,
${\overrightarrow F_g}$ the gravitational
acceleration and $\nu$ is the coefficient of kinematic viscosity
$=\alpha c_s^2/\Omega_{kn}$. As we are using the height integrated
equations we get from (2) the radial and angular momentum conservation
equations
   \begin{equation}
V \frac{dV}{dR} - R\Omega^2 - 2\omega\Omega R- \omega^2 R = - R
\Omega^2_{kn} - \frac{1}{\rho} \frac{dp}{dR}
   \end{equation}
 and
   \begin{equation}
V \frac{d}{dR} \left( \Omega R^2 \right) = \frac{1}{\rho RH} \frac{d}{dR}
\left(\nu \rho R^3 H \frac{d\Omega}{dR} \right) - 2\omega RV
   \end{equation}
with $V$ and $\Omega$ being the radial velocity and angular frequency
respectively of the fluid element.
 
We shall use the quasi-Newtonian potential
   \begin{equation}
\phi=GM/(R-R_g), \hskip.3cm R_g=2GM/c^2
   \end{equation}
in terms of which the Keplerian angular velocity is
   \begin{equation}
\Omega_{kn}^2=GM/R(R-R_g)^2.
   \end{equation}
 
As mentioned earlier we shall now couple $\omega$ and $\Omega$ through a
simple relation $\omega=a\Omega$, $a$ being a constant. Using the equation
of state $p=\rho c_s^2$ we can rewrite (3) and (4) as
 \begin{equation}
V\frac{dV}{dR}-nR\Omega^2+\Omega_{kn}^2 R +2 c_s \frac{dc_s}{dR}+
\frac{c_s^2}{\rho}\frac{d\rho}{dR}=0
   \end{equation}
with $n=(1+a)^2$ and
 \begin{equation}
\frac{d\Omega}{dR}=\frac{V\Omega_{kn}}{\alpha c_s^2 R^2}\left[\Omega R^2
+a\int2\Omega R dR-j\right]
 \end{equation}
where the constant $j$ has the usual interpretation of being the
angular momentum per unit accreted mass. However for self similar solutions
one assumes $j$ negligible in comparison with the other terms.
 
Apart from these hydrodynamical equations one also considers the thermodynamical
balance between the local viscous heating and radiative cooling that gives
rise to advection and is obtained through the energy equation\cite{ny}
 \begin{equation}
\frac{\rho V}{\gamma -1} \frac{d}{dR} c^2_s - Vc^2_s \frac{d\rho}{dR} =
\frac{f\alpha\rho c^2_s R^2}{\Omega_{kn}} \left( \frac{d\Omega}{dR}\right)^2.
 \end{equation}

\section{Equilibrium Solutions}
 
The original set of self similar solution were obtained by setting
$\frac{R_g}{R}=0$ in the potential, thus taking $\Omega_{kn}=\Omega_K=
(GM/R^3)^{1/2}$ and assuming the radial dependence of the other parameters to
be\\ $V=V_0 \left( \frac{R_g}{R}\right)^{1/2}$, $c_s=c_{s0} \left(\frac{R_g}{R}
\right)^{1/2}$,\\
$\Omega=\Omega_0 \left( \frac{R_g}{R}\right)^{3/2}$,
$\rho=\rho_0 \left( \frac{R_g}{R}\right)^{3/2}$.\\
Assuming the same dependences and substituting these in Eqns. (1), (7), (8) \&
(9), one can get the set of algebraic equations
 \begin{eqnarray}
V_0^2-1+2n\Omega_0^2R_g^2+5c_{s0}^2=0\nonumber \\
\frac{3}{2}c_{s0}^2=-V_0(1+4a)/\sqrt{2}\alpha \\
(3\gamma-5) V_0=\frac{9}{\sqrt{2}}(\gamma-1)f\alpha \Omega_0^2 R_g^2 \nonumber
\\
\rho_0=-{\dot M}/(4\pi\sqrt{5}c_{s0}V_0R_g^2).\nonumber
 \end{eqnarray}
Solving these, one can get
 \begin{eqnarray}
V_0=-Ag/(3\sqrt{2}\alpha),\nonumber \\
c_{s0}^2=Ag(1+4a)/9\alpha^2 \\
\Omega_0 R_g=({\epsilon^\prime} A g)^{1/2}/3\alpha, \nonumber \\
\rho_0=\frac{63}{88}\alpha^2\left[\frac{5}{2}(Ag)^3(1+4a)\right]^{-1/2},
\nonumber
 \end{eqnarray}
with
 \begin{eqnarray}
\epsilon^\prime=\frac{\left(\frac{5}{3}-\gamma\right)}{(\gamma-1)f}\nonumber \\
A=2n\epsilon^\prime+5(1+4a) \\
g=-1+\sqrt{1+\frac{18\alpha^2}{A^2}}. \nonumber
 \end{eqnarray}
Figures 1 and 2 show the behaviour of the various physical parameters for
the cases $a\ge0$ and $a\le0$ respectively. The behaviours are shown as a
function of $R$ for fixed values of $\alpha$ and $\gamma$. The general trend
of the solutions are similar for all values of $a$ (to be expected) except
for the fact that with $a\ne0$ the relative values tend to decrease with
higher $a$ values for the fluid velocities ( Fig. 1ab) in the co-rotating case,
but increase for the counter-rotating flow (Fig. 2ab). Whereas the sound speed
shows no variation for the co-rotating case (Fig. 1c) but very small change
in the counter-rotating case (Fig. 2c), the density shows an increase with the
values of $a$ for the co-rotating flow (Fig. 1d) and decrease in the
counter-rotating cases (Fig. 2d), the change being pronounced only
close to the compact object.

\begin{figure}
\vbox{
\vskip 1.0cm
\hskip 0.0cm
\centerline{
\psfig{figure=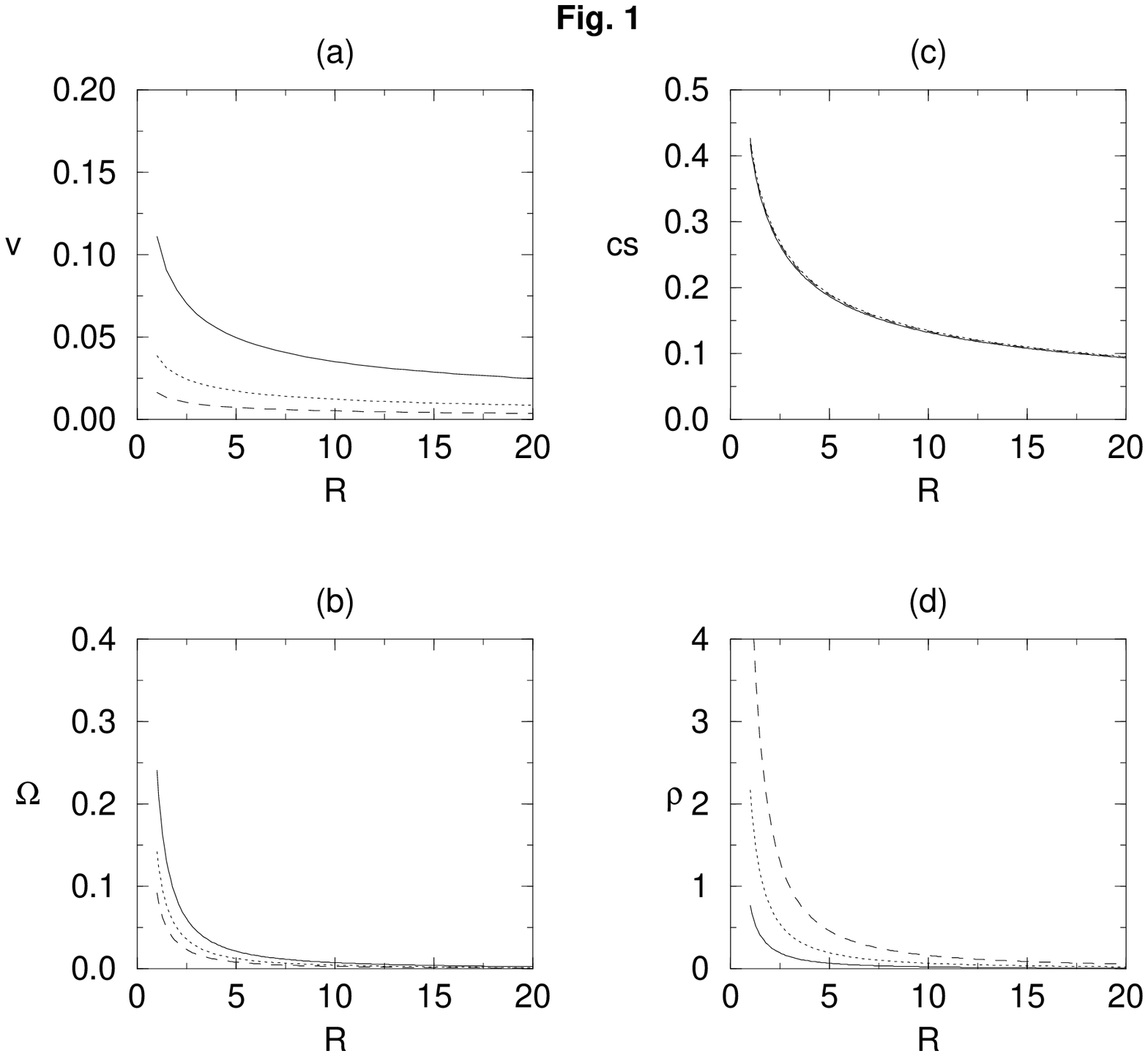,height=10truecm,width=10truecm,angle=0}}}
\vspace{-0.0cm}
\noindent {\small {\bf Fig. 1} : Plots of radial velocity $V$, angular
velocity $\Omega$, sound speed $c_s$ and density $\rho$ as a function of
$R$ for different values of the rotation parameter `$a$' for fixed
$\alpha=0.3$ and $\gamma=1.5$;  $a = 0$  ({\bf --------}), $a =0.5
\; (........)$ and $a = 1.5 \; \left( -\;\;-\;\;-\;\;-\;\;-\right)$.
}
\end{figure}
\begin{figure}
\vbox{
\vskip 1.0cm
\hskip 0.0cm
\centerline{
\psfig{figure=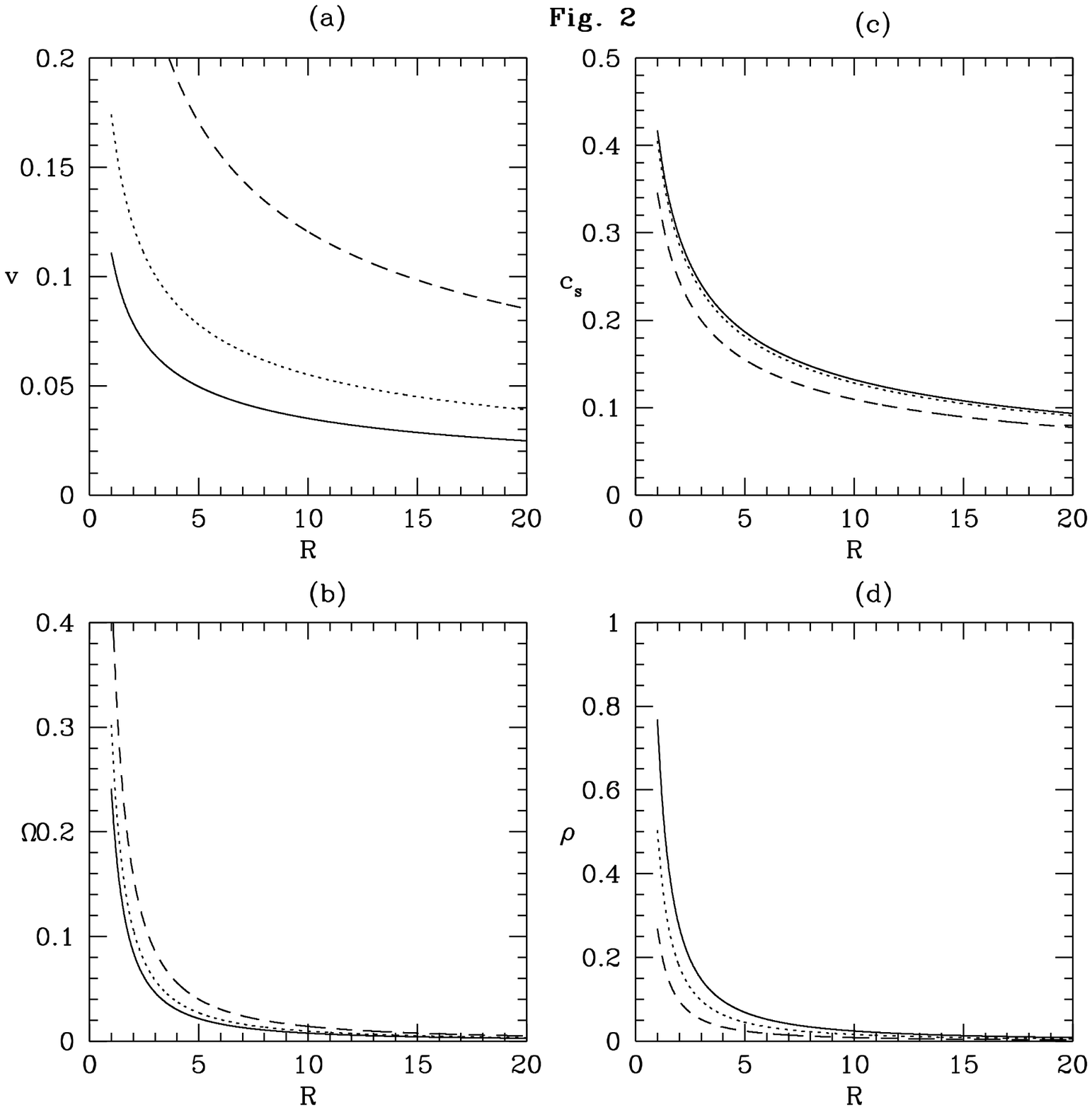,height=10truecm,width=10truecm,angle=0}}}
\vspace{-0.0cm}
\noindent {\small {\bf Fig. 2} : Plots similar as Fig. 1 but for
 $a = 0$  ({\bf --------}), $a =-0.1
\; (........)$ and $a = -0.2 \; \left( -\;\;-\;\;-\;\;-\;\;-\right)$.
}
\end{figure}
 
\section{ Perturbation}
Following the approach of Bhatt \& Prasanna\cite{bp} we now study the
effects of non-Newtonian potential on the self similar solutions considered
above, through perturbations \\
$p\rightarrow p_0+\delta p$, $V\rightarrow V_0+\delta V$,
$\Omega\rightarrow \Omega_0+\delta \Omega$, $c_s\rightarrow c_{s0}+\delta c_s$\\
in the equations of motion and mass and energy conservation equations.
We then simplify by retaining the terms linear in the perturbations, after
using the background solution (11) and obtain the equations governing
the perturbations. Perturbation in the Keplerian velocity to the first order
in $\frac{R_g}{R}$ is given by\\
$\delta \Omega_{kn}^2\approx 2 \frac{R_g}{R}\Omega_K^2$, $\delta \Omega_{kn}=
\frac{R_g}{R}\Omega_K$\\
where $\Omega_K$ is taken as $\left(GM/R^3\right)^{1/2}$.
 
Thus the set of equations governing the perturbations\\
$\delta V$, $\delta \Omega$, $\delta c_s$ and $\delta \rho$
are:
\footnote{There are a couple of minor typographical errors
in \cite{bp}, which can be corrected by taking these
equation with a=0.}
$$
\frac{\delta H}{H}+\frac{\delta\rho}{\rho}+\frac{\delta V}{V}=0;\hskip0.3cm
\frac{\delta H}{H}=\frac{\delta c_s}{c_s}-\frac{R_g}{R},
$$
$$
V \frac{d}{dR}\delta V + \delta V \frac{dV}{dR}= 2nR\Omega
\delta\Omega - R \delta \Omega^2_{kn}
+ \frac{\delta \rho}{\rho^2} \frac{dp}{dR} - \frac{1}{\rho} \frac{d}{dR}
\delta p,
$$
$$
\delta p = c^2_s \delta \rho + 2\rho c_s \delta c_s,
\eqno{(13)}
$$
$$
\alpha c^2_s \frac{d}{dR} \delta\Omega + 2\alpha c_s \frac{d\Omega}{dR}
\delta c_s= \left( \Omega + \frac{2a}{R^2} \int \Omega R dR \right)
\left( \Omega_{kn} \delta V + V \delta \Omega_{kn} \right)
+ V\Omega_{kn} \left( \delta \Omega + \frac{2a}{R^2} \int\delta\Omega RdR \right),
$$
$$
\frac{\rho V}{\gamma -1} \frac{dc^2_s}{dR} \left[ \frac{\delta
\rho}{\rho} + \frac{\delta V}{V} + 2 \left( \frac{dc^2_s}{dR}\right)^{-1}
\frac{d}{dR} \left( c_s \delta c_s \right) \right]- c^2_sV
\frac{d\rho}{dR} \left[ 2 \frac{\delta c_s}{c_s} + \frac{\delta V}{V} +
\left(\frac{d\rho}{dR}\right)^{-1} \frac{d}{dR} \delta\rho \right]
$$
$$
=\frac{f\alpha R^2c^2_s\rho}{\Omega_k} \left( \frac{d\Omega}{dR}\right)^2
\left[ \frac{\delta \rho}{\rho} + \frac{2\delta c_s}{c_s} - \frac{R_g}{R} +
2\left(\frac{d\Omega}{dR}\right)^{-1}\frac{d}{dR}\delta\Omega\right].
$$
Here again as in the equilibrium case, for self similar solutions, the
assumption $j<<\delta(\Omega R^2 + 2a\int\Omega R dR)$ is maintained.
Thus again using self similarity of the perturbed solutions through
\setcounter{equation}{13}
\begin{eqnarray}
\delta V=V_{01}\left(\frac{R_g}{R}\right)^{3/2}, \hskip0.1cm
\delta c_s=c_{s01}\left(\frac{R_g}{R}\right)^{3/2},\hskip0.1cm
\delta \Omega=\Omega_{01}\left(\frac{R_g}{R}\right)^{5/2},\hskip0.1cm
\delta \rho=\rho_{01}\left(\frac{R_g}{R}\right)^{5/2}
\end{eqnarray}
and substituting these in (13) one gets the set of algebraic equations:\\
 
\begin{equation}
\frac{\rho_{01}}{\rho_0} + \frac{V_{01}}{V_0} + \frac{c_{s01}}{c_{s0}} = 1,
\end{equation}
\begin{equation}
\frac{Ag}{9\alpha^2} \left( Ag - \left( 1 + 4a\right)\right)
\frac{V_{01}}{V_0} + \frac{2n\epsilon^\prime Ag}{9\alpha^2}
\frac{\Omega_{01}}{\Omega_0} +
\frac{6Ag(1+4a)}{9\alpha^2} \frac{c_{s01}}{c_{s0}} - 1
+\frac{Ag(1+4a)}{9\alpha^2} = 0,
\end{equation}
\begin{equation}
-\frac{3}{2} \frac{V_{01}}{V_0} + \left( \frac{5}{2} - \frac{3}{2}
\frac{(1-4a)}{(1+4a)} \right) \frac{\Omega_{01}}{\Omega_0} +
3\frac{c_{s01}}{c_{s0}} =  \frac{3}{2},
\end{equation}
\begin{equation}
\left(\gamma - 3 \right) \frac{V_{01}}{V_0} + 10 \left( \frac{5}{3} -
\gamma \right)\frac{\Omega_{01}}{\Omega_0} - 2 \left( \gamma + 1 \right)
\frac{c_{s01}}{c_{s0}} = 7  - 5 \gamma.
\end{equation}
Since it is a consistent set of inhomogeneous equations, one can solve for
the perturbations and thus get the set of solutions incorporating the
quasi Newtonian potential, to be
\begin{eqnarray}
\rho_p=\rho_0\left(\frac{R_g}{R}\right)^{3/2}\left[1+\frac{\rho_{01}}{\rho_0}
\frac{R_g}{R}\right],\nonumber\\
V_p=V_0\left(\frac{R_g}{R}\right)^{1/2}\left[1+\frac{V_{01}}{V_0}
\frac{R_g}{R}\right],\\
c_{sp}=c_{s0}\left(\frac{R_g}{R}\right)^{1/2}\left[1+\frac{c_{s01}}{c_{s0}}
\frac{R_g}{R}\right],\nonumber\\
\Omega_{p}=\Omega_{0}\left(\frac{R_g}{R}\right)^{3/2}
\left[1+\frac{\Omega_{01}}{\Omega_{0}}\frac{R_g}{R}\right].\nonumber
\end{eqnarray}

\begin{figure}
\vbox{
\vskip 1.0cm
\hskip 0.0cm
\centerline{
\psfig{figure=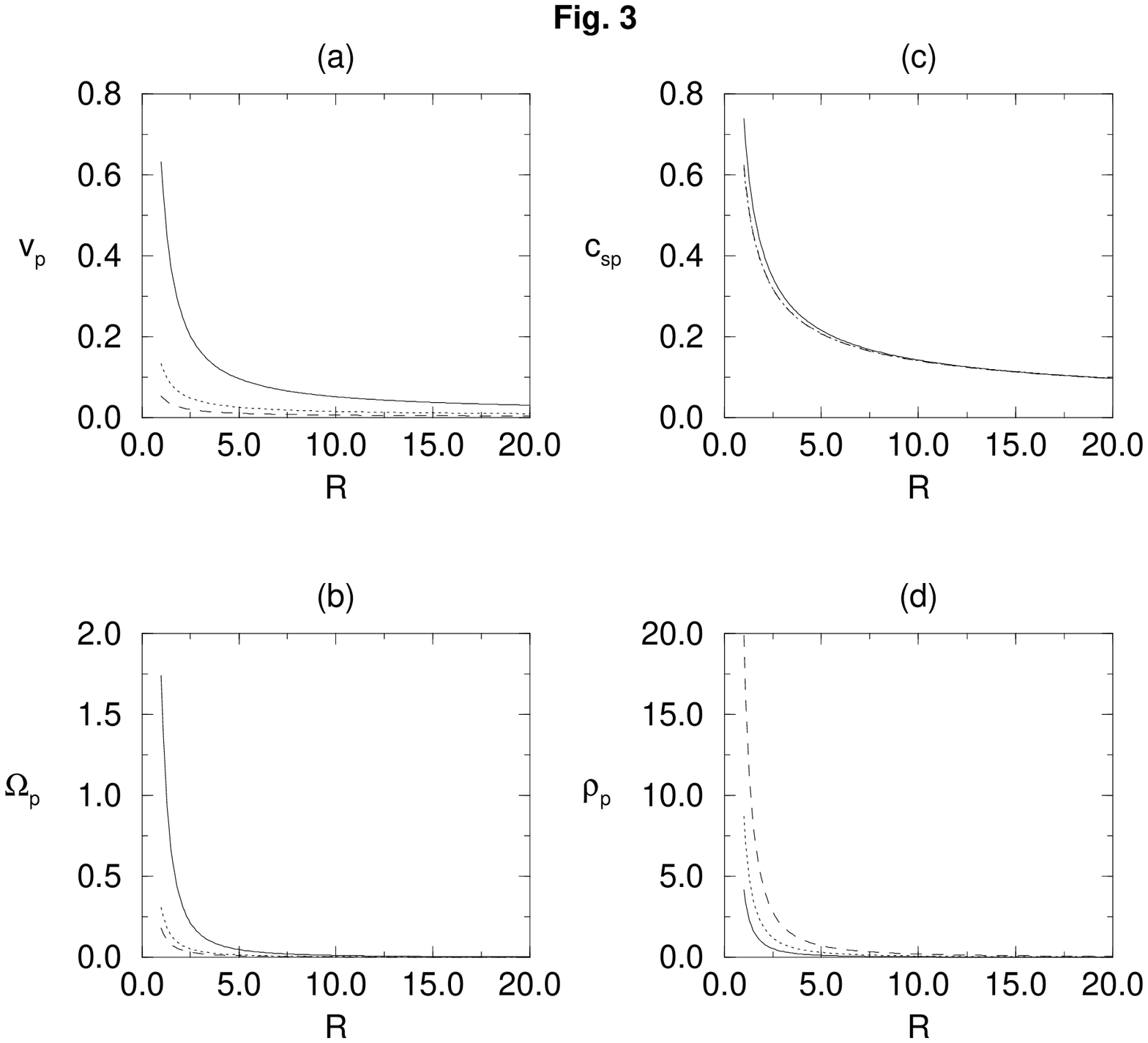,height=10truecm,width=10truecm,angle=0}}}
\vspace{-0.0cm}
\noindent {\small {\bf Fig. 3} :
Same as Fig. 1 but with the addition of perturbative
solutions arising from quasi-Newtonian potential;
 $a = 0$  ({\bf --------}), $a =0.5
\; (........)$ and $a = 1.5 \; \left( -\;\;-\;\;-\;\;-\;\;-\right)$.
}
\end{figure}

Figures (3)-(5) present these solutions for the physical quantities as a
function of $R$ with two of the three parameters $a$, $\gamma$, $\alpha$
fixed, while the other is varied. While the changes as a function of $a$
remains almost the same as in the non-perturbed case (Fig. 3), there
appear some changes as a function of the viscosity ($\alpha$) and
gas parameter ($\gamma$). As may be noted in Fig. 4a the radial velocity
shows a jump in the profile as $\gamma$ changes from $1.1$ to $1.6$,
while the other three do not show such a jump (Fig. 4b,c,d). On the
other hand as is depicted in Fig. 5, their behaviour as a function of
$\alpha$ for fixed $\gamma=1.5$ and $a=0.5$ is quite interesting. While
$V_p$ increases with $\alpha$ (Fig. 5a) and $\rho_p$ decreases
(Fig. 5d), $\Omega_p$ and $c_p$ show no change at all with varying
$\alpha$ (Fig. 5b,c).
 
\begin{figure}
\vbox{
\vskip 1.0cm
\hskip 0.0cm
\centerline{
\psfig{figure=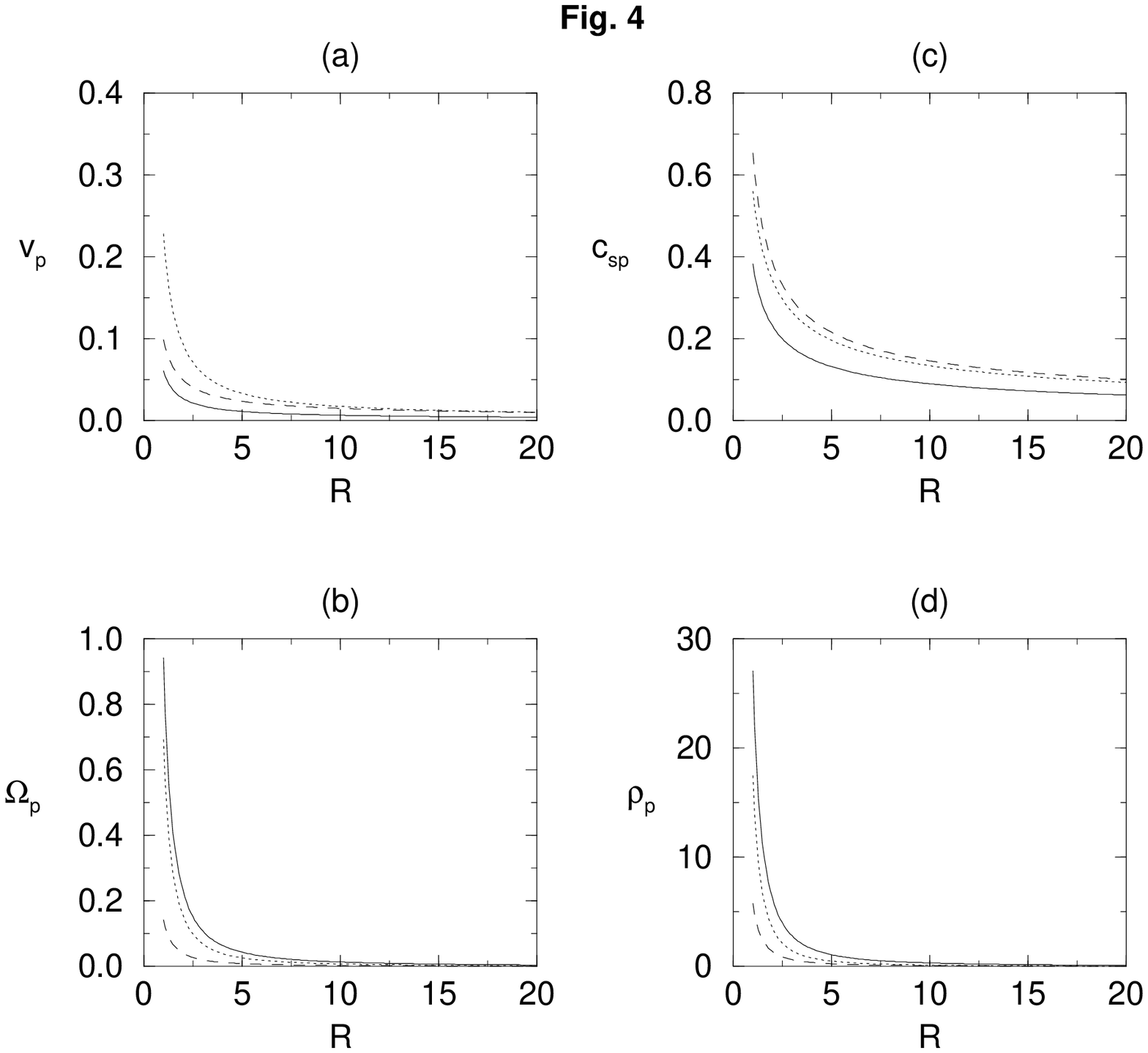,height=10truecm,width=10truecm,angle=0}}}
\vspace{-0.0cm}
\noindent {\small {\bf Fig. 4} :
Same as Fig. 3 for fixed $\alpha=0.3$ and $a=0.5$,
but with different $\gamma$s; $\gamma=1.1$ ({\bf --------}),
$\gamma=1.4 \; (........)$ and $\gamma=1.6 \;
\left( -\;\;-\;\;-\;\;-\;\;-\right)$.
}
\end{figure}
\begin{figure}
\vbox{
\vskip 1.0cm
\hskip 0.0cm
\centerline{
\psfig{figure=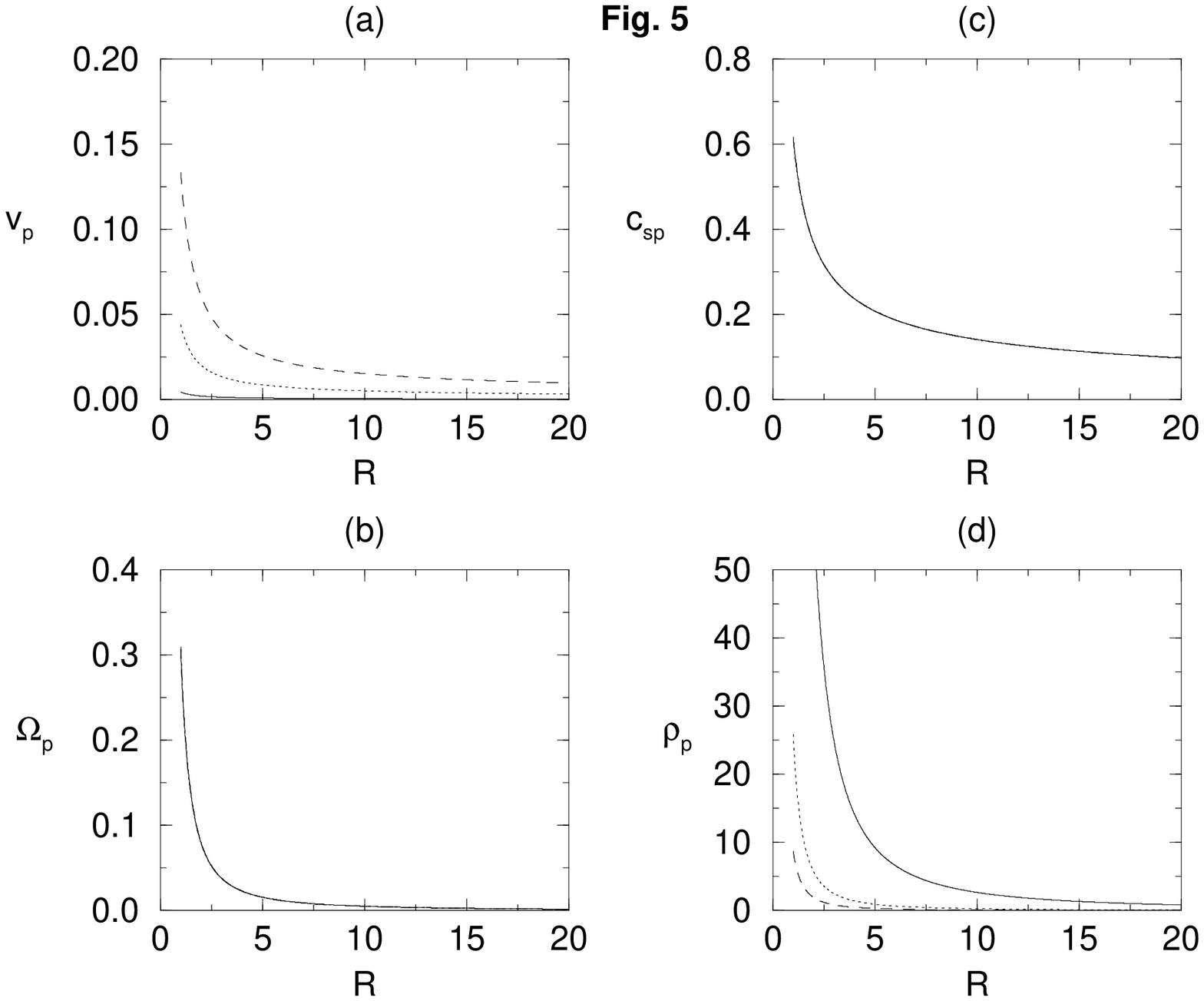,height=10truecm,width=10truecm,angle=0}}}
\vspace{-0.0cm}
\noindent {\small {\bf Fig. 5} :
Same as Fig. 3 for fixed $\gamma=1.5$ and $a=0.5$,
but varying $\alpha$; $\alpha=0.01$ ({\bf --------}),
$\alpha=0.1 \; (........)$ and $\alpha=0.3 \;
\left( -\;\;-\;\;-\;\;-\;\;-\right)$.
}
\end{figure}
\begin{figure}
\vbox{
\vskip 1.0cm
\hskip 0.0cm
\centerline{
\psfig{figure=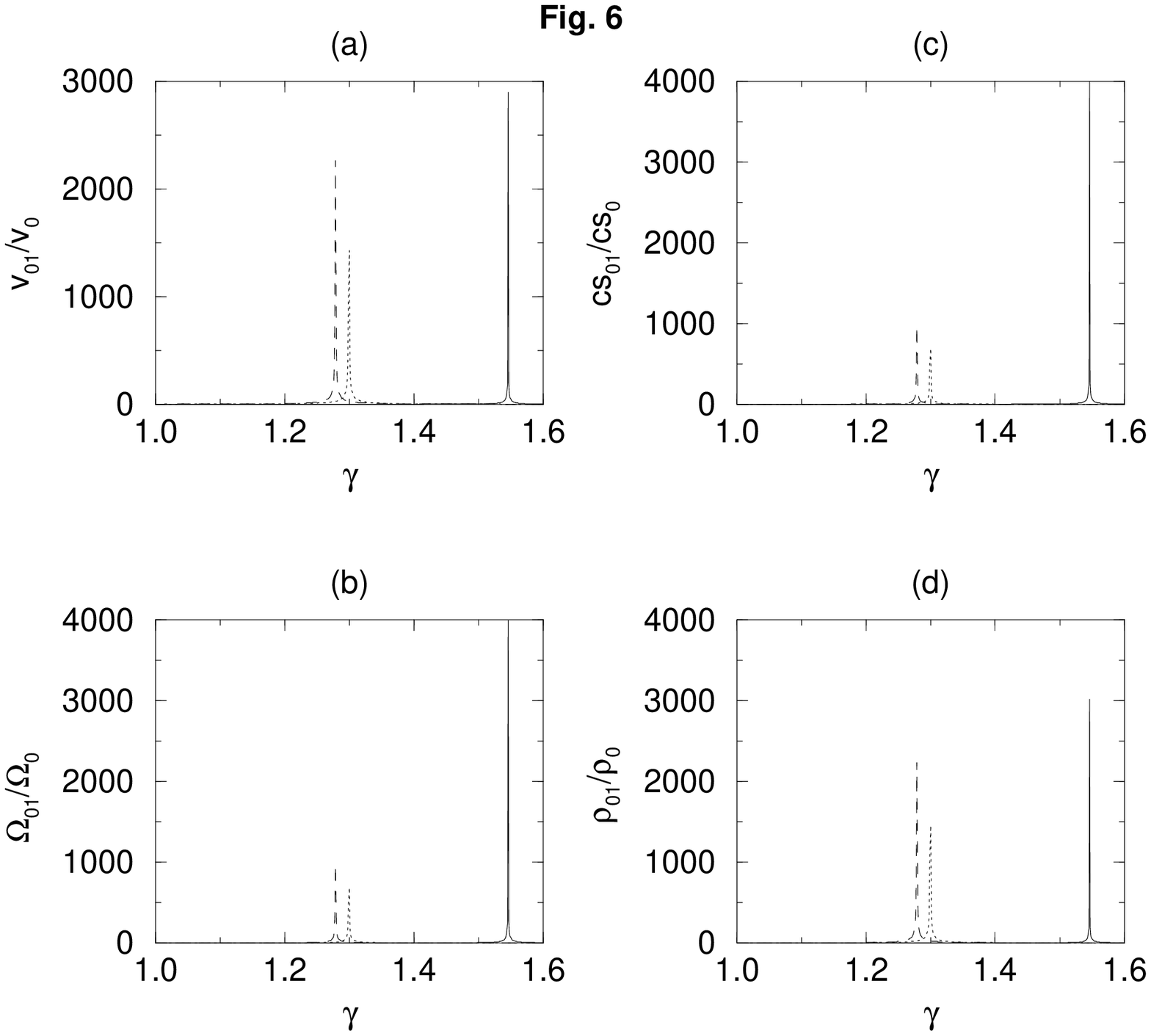,height=10truecm,width=10truecm,angle=0}}}
\vspace{-0.0cm}
\noindent {\small {\bf Fig. 6} :
Plots of the ratios of perturbed to the unperturbed
parameters as a function $\gamma$, for fixed $\alpha=0.3$ and different
values of $a$;  $a=0$ ({\bf --------}),
$a=0.5 \; (........)$ and $a=1.5 \; \left( -\;\;-\;\;-\;\;-\;\;-\right)$.
}
\end{figure}

In order to understand these features, we have presented the plots of the
ratios of perturbed to the unperturbed quantities ($V_{01}/V_0$,
$\Omega_{01}/\Omega_0$, $c_{s01}/c_{s0}$, $\rho_{01}/\rho_{0}$) in
Figs. (6) and (7) one as a function of $\gamma$ and the other as a
function of $\alpha$ for the cases $a=0,0.5,1.5$. As is evident from
Fig. 6 the singularity in the perturbations has shifted to lower values
of $\gamma$ as $a$ changes from $0$ to $1.5$. For $a=0$, the location of the

\begin{figure}
\vbox{
\vskip 1.0cm
\hskip 0.0cm
\centerline{
\psfig{figure=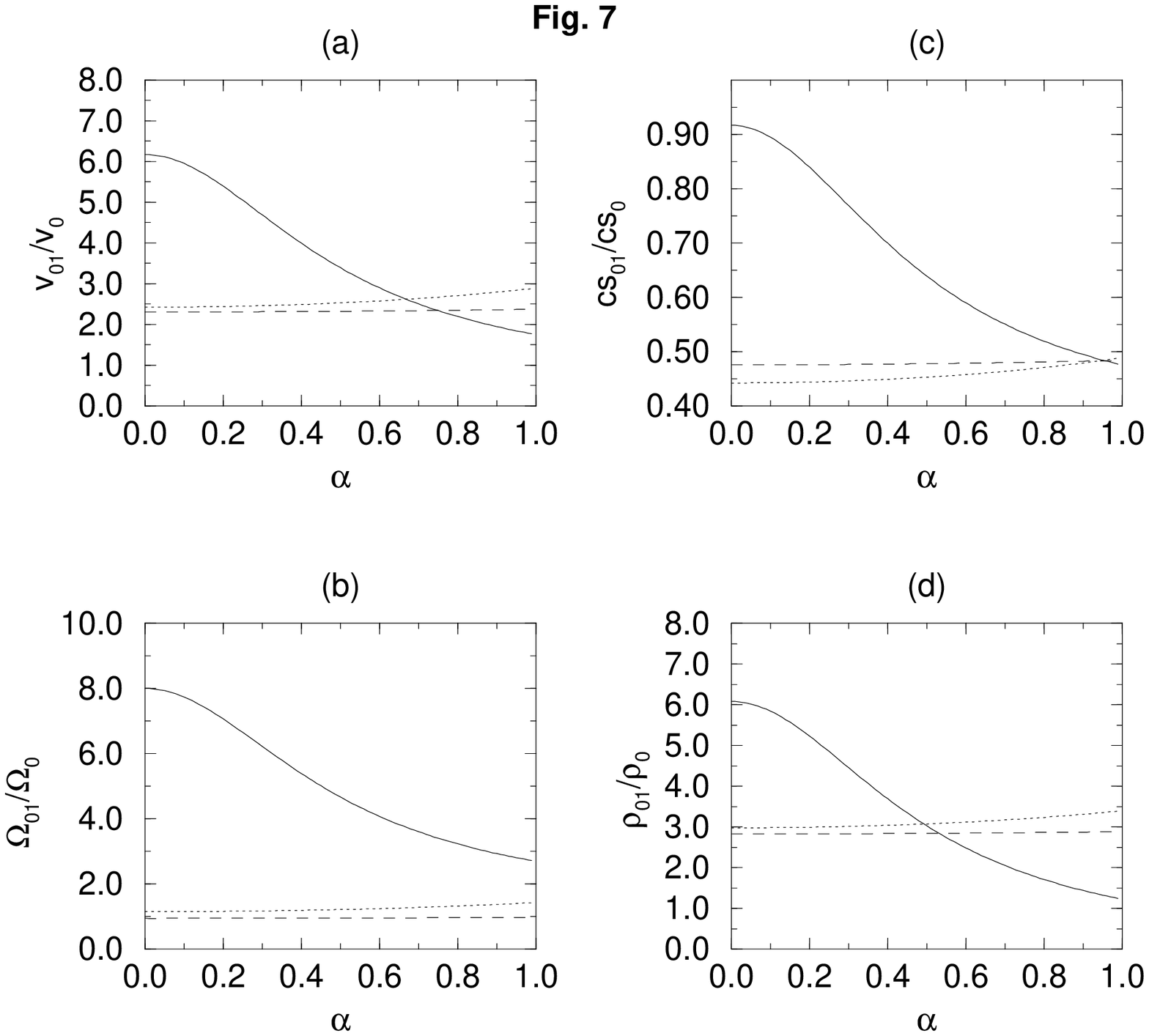,height=10truecm,width=10truecm,angle=0}}}
\vspace{-0.0cm}
\noindent {\small {\bf Fig. 7} :
Plots same as in Fig. 6 for fixed $\gamma=1.5$ as a
function of $\alpha$ and different values of $a$;  $a=0$ ({\bf --------}),
$a=0.5 \; (........)$ and $a=1.5 \; \left( -\;\;-\;\;-\;\;-\;\;-\right)$.
}
\end{figure}

singularity in the $\gamma$ space is at $\gamma=1.55$, a result which was
obtained earlier by Bhatt \& Prasanna\cite{bp}. Figure 7 shows the perturbation
in the $\alpha$-space, whereas for $a=0$ the perturbations grows as the
flow gets nearer to the compact object, (a result known earlier), with the
inclusion of Coriolis terms $a\ne0$, the perturbations are well confined
and further the ratio decreases as the flow approaches the central gravitating
source. Fig. 8 depicts the plots of the ratios of the perturbed to
unperturbed physical quantities for $a=0, -0.1, -0.2$ as a function of
$\gamma$ for fixed $\alpha=0.3$. It can be clearly seen that there
appears no singular behaviour for $a\ne0$ unlike the case $a=0$.

\begin{figure}
\vbox{
\vskip 1.0cm
\hskip 0.0cm
\centerline{
\psfig{figure=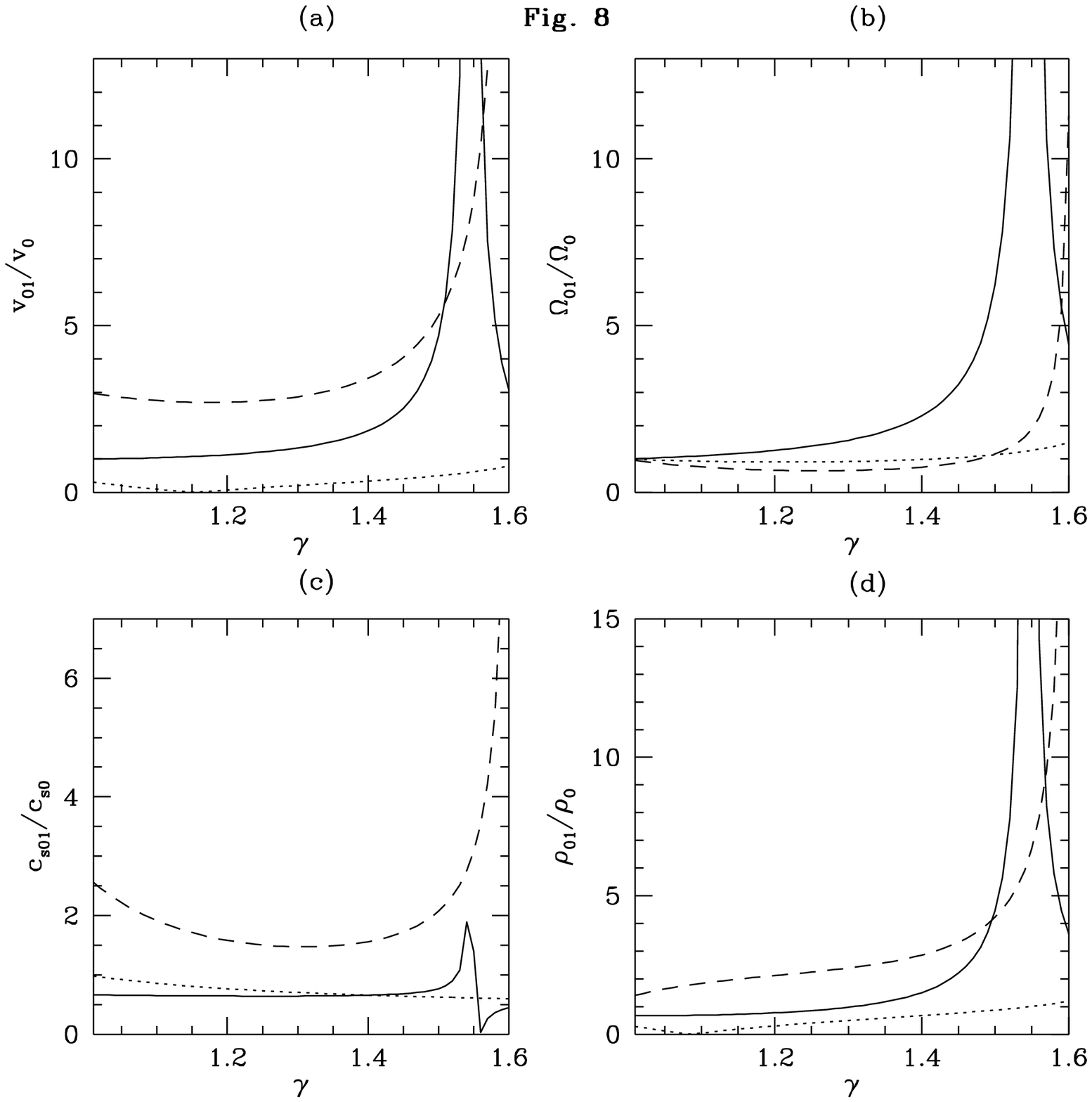,height=10truecm,width=10truecm,angle=0}}}
\vspace{-0.0cm}
\noindent {\small {\bf Fig. 8} :
Plots similar as Fig. 6 but for $a$ negative;
$a=0$ ({\bf --------}), $a=-0.1 \; (........)$ and
$a=-0.2 \; \left( -\;\;-\;\;-\;\;-\;\;-\right)$.
}
\end{figure}

\section{Discussion}

The introduction of the Coriolis force and a direct coupling of the two
angular velocities seem to bring in some features which could be helpful
for understanding the parameter space available for model building.
At the outset one can see that the reality of the solutions, immediately
requires $(1+4a)>0$, meaning that the counter-rotating flows have to have
a minimum speed, at least greater than four times the angular velocity
of the central source, to be in equilibrium. This may not be very surprising,
as the influence of the central rotation of `frame dragging' would imply that
any counter-rotating body has to overcome this barrier. Though in
principle `frame dragging' is purely a general relativistic effect, it
appears as though the Coriolis term in Newtonian physics mimics a similar effect.
An important outcome of the analysis is that the rotational effect seems
to stabilise the flows for the usually considered gas and viscosity parameters
$\gamma$ and $\alpha$ respectively as depicted by the perturbations of
the self similar solution; ($4/3\le\gamma\le 5/3$, $0<\alpha\le0.3$).
 
\begin{figure}
\vbox{
\vskip 1.0cm
\hskip 0.0cm
\centerline{
\psfig{figure=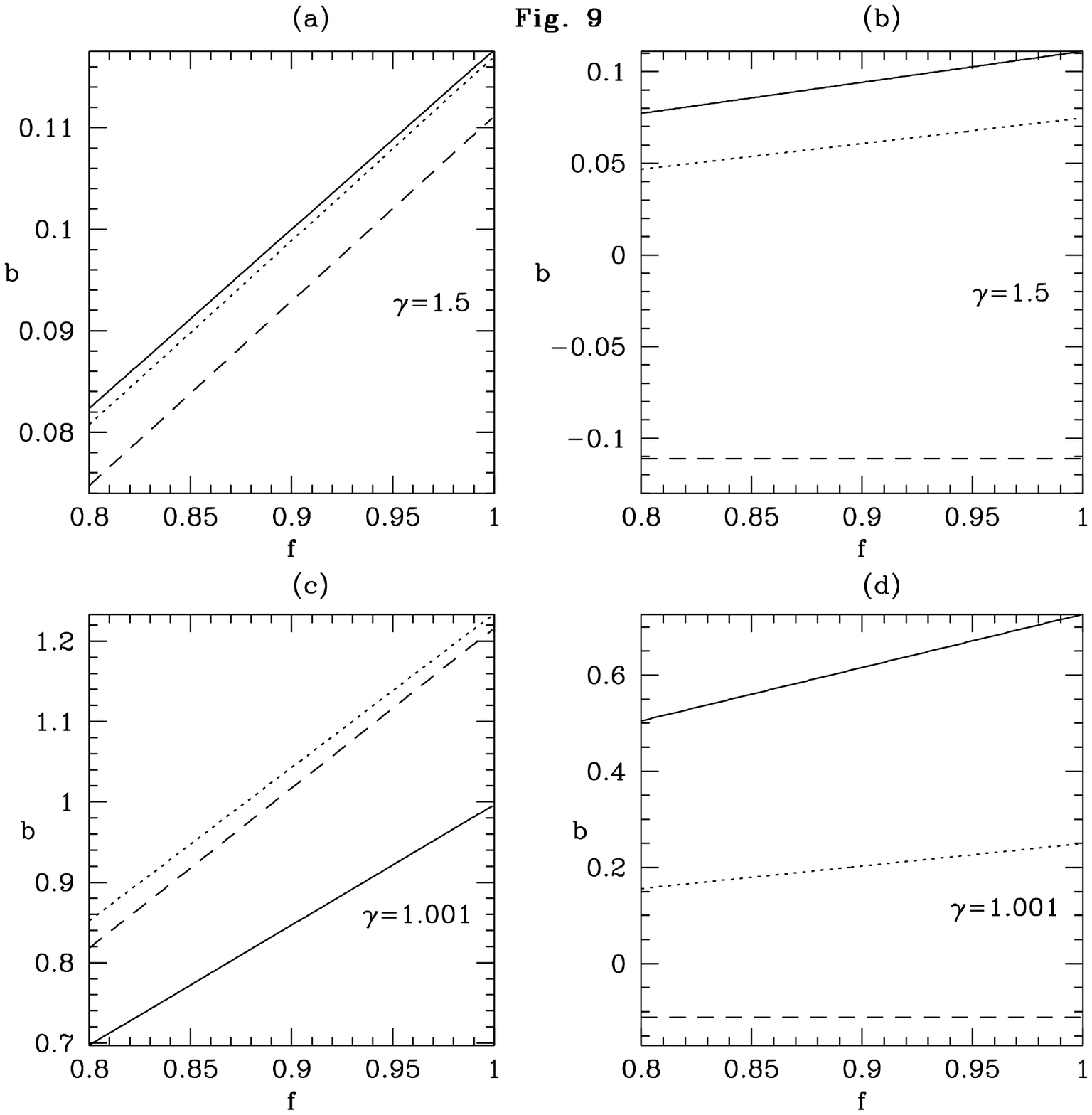,height=10truecm,width=10truecm,angle=0}}}
\vspace{-0.0cm}
\noindent {\small {\bf Fig. 9} :
Bernoulli parameter $b$ as a function of $f$ for:
(a),(c) $a=0$ ({\bf --------}), $a=0.25 \; (........)$ and
$a=0.5 \; \left( -\;\;-\;\;-\;\;-\;\;-\right)$;
(b),(d) $a=-0.1$ ({\bf --------}), $a=-0.2 \; (........)$ and
$a=-0.25 \; \left( -\;\;-\;\;-\;\;-\;\;-\right)$.
}
\end{figure}
\begin{figure}
\vbox{
\vskip 1.0cm
\hskip 0.0cm
\centerline{
\psfig{figure=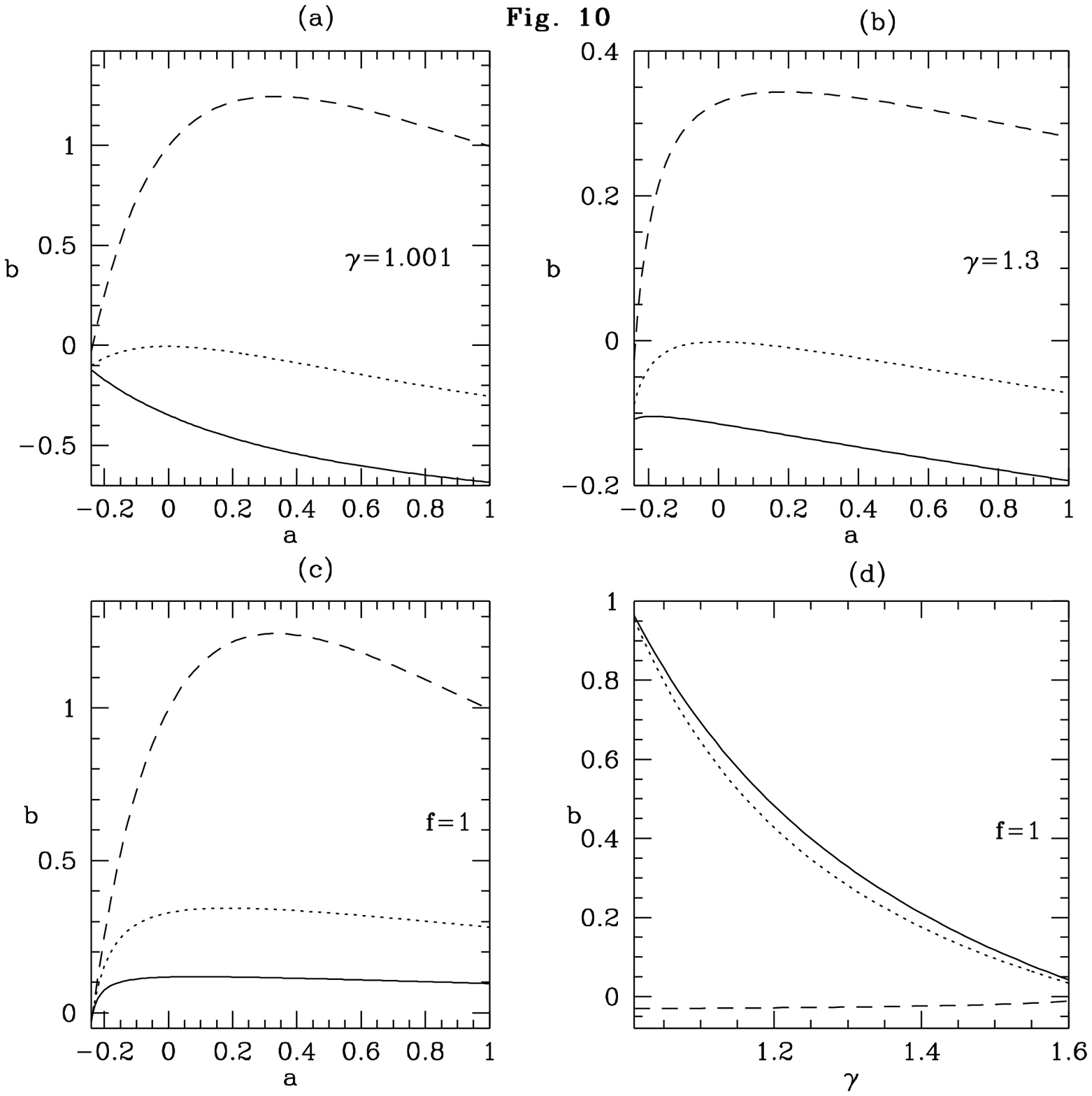,height=10truecm,width=10truecm,angle=0}}}
\vspace{-0.0cm}
\noindent {\small {\bf Fig. 10} :
Plots of $b$:
(a),(b) as a function of $a$; $f=0.1$ ({\bf --------}), $f=0.33 \; (........)$
and
$f=1 \; \left( -\;\;-\;\;-\;\;-\;\;-\right)$;
(c) as a function of $a$; $\gamma=1.5$ ({\bf --------}),
$\gamma=1.3 \; (........)$ and
$\gamma=1.001 \; \left( -\;\;-\;\;-\;\;-\;\;-\right)$;
(d) as a function of $\gamma$; $a=0$ ({\bf --------}), $a=1 \; (........)$ and
$a=-0.24 \; \left( -\;\;-\;\;-\;\;-\;\;-\right)$.
}
\end{figure}

Equation (7) shows that the solution satisfies the relation
\begin{equation}
\frac{V^2}{2} + n \Omega^2R^2 - \Omega^2_K R^2 + \frac{5}{2} c^2_s = 0.
\end{equation}
Using this in the computation of the normalised  Bernoulli parameter
`$b$' as defined by Narayan \& Yi \cite{ny}
\begin{equation}
b=\frac{1}{V^2_K} \left( \frac{V^2}{2} + \frac{\Omega^2R^2}{2} -
\Omega^2_K R^2 + \frac{\gamma}{\gamma - 1} c^2_s \right)
\end{equation}
one finds after using Eqns. (11) and (12)
\begin{equation}
b = \frac{Ag \epsilon^\prime}{19\alpha^2} \left[ 1 - 2n + 3f \left( 1 +
4a \right) \right]
\end{equation}
\begin{equation}
b\approx \left[\frac{\left( 1 - 2n + 3f \left( 1 + 4a
\right)\right)}{\left( 2n + \frac{5f}{\epsilon}
\left( 1 + 4a \right) \right)}\right].
\end{equation}
Figures (9) and (10) show the behaviour of `$b$' for different
ranges of parameters $\alpha$, $\gamma$, $f$. For $1<\gamma<\frac{5}{3}$,
$b$ is positive for $f\ge1/3$, for $a<-0.1$ and as $a$ approaches $-0.2$,
$b$ tends to become negative even for values of $f>1/3$. As $a$ tends
to the value $-0.25$, $b$ is negative for all values of $f$ and $\gamma$.
Thus while the co-rotating fluid can have energy transfer only outwards
for $f>1/3$, the counter-rotating fluid can have it in either direction
depending upon its angular velocity as compared to that of the central
accreting source. For advection dominated flows ($f=1$) $b$ is mostly
positive for $a>-0.2$. However $b$ changes sign for co-rotating flows
at $a=2+\sqrt{5}$ and for counter-rotating flows at $a=2-\sqrt{5}$. Hence
in principle if the energy transfer inwards has to be effective, the
co-rotating flow has to have very low angular velocity $\Omega<\omega/(2
+\sqrt{5})$, whereas the counter-rotating flow has to have very large
angular velocity $\Omega >\omega/(2-\sqrt{5})$.
 
In conclusion it may be seen that introducing rotational effects into a
Newtonian description of the accretion flow, through the Coriolis term, has
indeed enlarged the parameter space of the self similar solutions,
making them viable for model building scenarios, particularly treating
both co and counter rotating flows. A simple coupling of the angular
velocities has distinctly shown the possibility of energy transfer
both inwards and outwards depending on the effective angular
velocity of the fluid flow. More detailed studies, like looking for
a global solution and inclusion of other effects like convection,
may perhaps reveal interesting features which could influence the
`energy budget' of accretion dynamics.

\section*{Acknowledgments}
It is a pleasure to thank Ewald M\"uller and Uli Anzer for helpful
discussions and J. Banerji (PRL) for help in preparing the plots.
The hospitality (for ARP) at the Max-Planck-Institut f\"ur Astrophysik
where part of the work was carried out is gratefully acknowledged.

\end{document}